\documentclass[%
twocolumn,
floatfix,
amsmath,
amssymb,
aps,
notitlepage
]{revtex4-1}

\usepackage[english]{babel}

\usepackage[hidelinks=true]{hyperref}
\hypersetup{
    colorlinks,
    linkcolor={blue!80!black},
    citecolor={blue!80!black},
    urlcolor={blue!80!black}
}

\usepackage{setspace}
\usepackage{caption}
\usepackage{siunitx}
\usepackage{subcaption}
\usepackage{graphicx}
\usepackage{dcolumn}
\usepackage{bm}
\usepackage{gensymb}

\usepackage{xcolor}
\pagecolor{white}

\newcommand{\re}[1]{\textcolor{black}{#1}}

\begin{document}

\title{Rheology of periodically sheared suspensions undergoing \\reversible-irreversible transition}

\author{Zhouyang Ge$^{1,2}$}
\author{Gwynn J.~Elfring$^1$}

\affiliation{$^1$Department of Mechanical Engineering and Institute of Applied Mathematics, 
University of British Columbia, Vancouver V6T 1Z4, BC, Canada}
\affiliation{$^2$Department of Engineering Mechanics, 
KTH Royal Institute of Technology, 100 44 Stockholm, Sweden}

\date{\today}

\begin{abstract}

\re{The rheology of non-colloidal suspensions under cyclic shear is studied numerically.
The main findings are a strain amplitude ($\gamma_0$) dependent response 
in the shear stress and second normal stress difference ($N_2$).
Specifically, we find
a reduced viscosity,
an enhanced intracycle shear thinning, 
the onset of a finite $N_2$ and 
its frequency doubling,
all near a critical strain amplitude $\gamma_c$ that scales with the volume fraction $\phi$ as $\gamma_c \sim \phi^{-2}$.
These rheological changes also signify a reversible-irreversible transition (RIT),
dividing stroboscopic particle dynamics into a reversible absorbing phase (for $\gamma_0<\gamma_c$)
and a persistently diffusing phase (for $\gamma_0>\gamma_c$).
We explain the results based on two flow-induced mechanisms 
and elucidate their connection in the context of RIT through the underlying microstructure,
which tends towards hyperuniformity near $\gamma_0=\gamma_c$.}
Overall, we expect this correspondence between rheology and emergent dynamics to hold in a wide range of settings
where structural organizations are dominated by volume exclusions.

\end{abstract}

\maketitle

\section{Introduction}

\re{When a concentrated suspension is driven by shear,
the suspending particles can undergo diffusive motion even in the absence of inertial or Brownian effects.
Although this has long been observed \cite{Eckstein1977} 
and was subsequently understood as shear-induced diffusion \cite{davis1996hydrodynamic},
recent studies on periodically sheared suspensions revealed a remarkable dynamical phase transition 
that can occur without complex hydrodynamic interactions \cite{Pine_Nature_2005, Corte_NatPhys_2008}.
This phenomenon, now called \emph{reversible-irreversible transition} (RIT) \cite{Menon_Ramaswamy2009}
or absorbing-phase transition,
refers to the change in particle dynamics viewed stroboscopically:
depending on the driving strain amplitude, 
the particle motions are either reversible (absorbing) or irreversible (diffusive),
with a transition characterized by a critical strain amplitude.
Because of the ubiquity of suspensions in nature and their increasing relevance in technology,
new understanding of suspension dynamics under cyclic shear
has since led to broad developments in 
the studies of 
jamming and yielding \cite{das2020unified, Ness_Cates_PRL2020, Wilken2021random},
of self-organization \cite{Corte_etal_PRL2009, franceschini2011transverse, Royer49}
and memory formation \cite{keim2011generic, paulsen2014multiple, Keim2019memory} in disordered systems,
as well as in the fabrication of hyperuniform materials \cite{Hexner2015, tjhung2015hyperuniform, Wilken2020}
with potential applications in photonics, among others \cite{torquato2018hyperuniform}.}

Unlike the clear connection between emergent dynamical and structural properties of periodically sheared suspensions,
less is known about the relation between these dynamics and the suspension \emph{rheology}.
Conventionally, rheological properties of complex fluids are characterized in
either simple shear flows (to obtain the nonlinear viscosity at different shear rates)
or oscillatory shear flows (to obtain the linear viscoelasticity at different oscillation frequencies).
For suspensions of non-colloidal spheres, in particular,
viscous effects usually dominate over elasticity \cite{mewis_wagner_book},
and previous experiments on such systems have found that the suspension viscosity in oscillatory shear can depend on 
either the strain amplitude \cite{breedveld2001shear, Bricker_Butler2006, Lin_Phan-Thien_Khoo_2013, Ness_OS_2017SM, Martone_experiment},
or both the strain amplitude and the oscillation frequency \cite{breedveld2001shear, narumi2005response, Martone_experiment}.
Despite the complex dependence,
rheological properties of a suspension are theoretically fully determined from its microstructure and the interactions among constituent particles.
The general understanding is that shearing induces anisotropies in the microstructure \cite{gadala1980shear, brady1997microstructure},
which further leads to non-Newtonian behaviors (e.g.~normal stress differences) under simple shear \cite{sierou2002rheology, morris2009review}, 
as well as strain-dependent asymmetric stress responses under shear reversal \cite{gadala1980shear}.
In addition, irreversible particle dynamics and rate-dependent suspension rheology
may arise from weak non-hydrodynamic interactions (e.g.~van der Waals forces),
where the effect is typically stronger at smaller strain amplitudes \cite{ge2021irreversibility}. 
Therefore, an indirect connection must exist between rheology and emergent dynamics for suspensions under cyclic shear,
as both are related to the underlying microstructure.
Although this might have been realized to some extent in the past \cite{breedveld2001shear, Corte_NatPhys_2008, Lin_Phan-Thien_Khoo_2013},
a clear physical picture of their relationship in the context of RIT has not been established so far.

In this paper, we identify generic rheological responses of periodically driven suspensions that are also undergoing RIT.
Specifically, we numerically show that non-colloidal suspensions under cyclic shear
have \re{at least four rheological signatures near their critical transition points:
(I) reduced relative viscosities,
(II) enhanced intracycle shear thinning, 
(III) the onset of a finite second normal stress difference and 
(IV) its peculiar frequency doubling within a cycle.}
In addition, the critical strain amplitude $\gamma_c$ over which the particle dynamics change most significantly
can be inferred from the above rheological measures and scales with the volume fraction $\phi$ as $\gamma_c \sim \phi^{-2}$.
The scaling suggests the importance of many-body couplings which, 
counterintuitively, do not require far-field hydrodynamic interactions. 
\re{We explain these results on the basis of the stress asymmetry caused by flow reversal \cite{gadala1980shear} 
and the phenomenon of shear-induced diffusion \cite{leighton_acrivos_1987}
that were known experimentally but not fully recognized in the context of RIT.}
Furthermore, microstructural analysis of the suspension near its critical point links these observations to a common configuration
characterized by a local anisotropy and suppression of long-range density fluctuations;
the latter further implies hyperuniformity \cite{torquato2003local, torquato2018hyperuniform}.
\re{Overall, our results highlight a nontrivial connection between rheology and collective dynamics in periodically sheared suspensions.
More generally, we believe rheology may serve as an indicator of both the underlying structure and the emergent dynamics.}

\section{Model}

We simulate particle suspensions in shear flow using a method called Fast Stokesian Dynamics (FSD) \cite{fiore2019fast}.
FSD is a recent version of the Stokesian Dynamics (SD) that is
commonly used for modeling of (non-)colloidal suspensions \cite{sd1988}.
The core of SD is a multipole expansion of the boundary integral equation for particles in Stokes flow,
linearly relating the hydrodynamic forces and torques ${\bm F}^\text{H}$ on the particles 
to their translational and angular velocities ${\bm U}$ through a resistance tensor ${\bm R}_\text{FU}$,
viz.~${\bm F}^\text{H}=-{\bm R}_\text{FU}\cdot {\bm U}$.
Assuming negligible particle inertia, 
${\bm F}^\text{H}$ is always in balance with the prescribed external force ${\bm F}^\text{Ext}$ on each particle,
viz.~${\bm F}^\text{H} + {\bm F}^\text{Ext} =0$;
thus, the above equations are closed provided ${\bm R}_{FU}$ and its inverse are known.
All SD methods decompose the hydrodynamic resistance into a far-field and a near-field part, 
viz.~${\bm R}_\text{FU}={\bm R}_\text{FU}^\text{ff} + {\bm R}_\text{FU}^\text{nf}$,
\re{allowing for independent control of the two contributions.
Specifically, the far-field interactions can be turned off by setting the off-diagonal terms of ${\bm R}_\text{FU}^\text{ff}$ to zero.
The same decomposition also applies to the stress calculation, 
which depends on both far-field and near-field hydrodynamic interactions
but does not affect the particle dynamics \cite{sd1988}.}
FSD is a fast algorithm to numerically solve the particle dynamics utilizing graphical processing units (GPU).

\begin{figure}
	\centering
	\includegraphics[height=2.8cm]{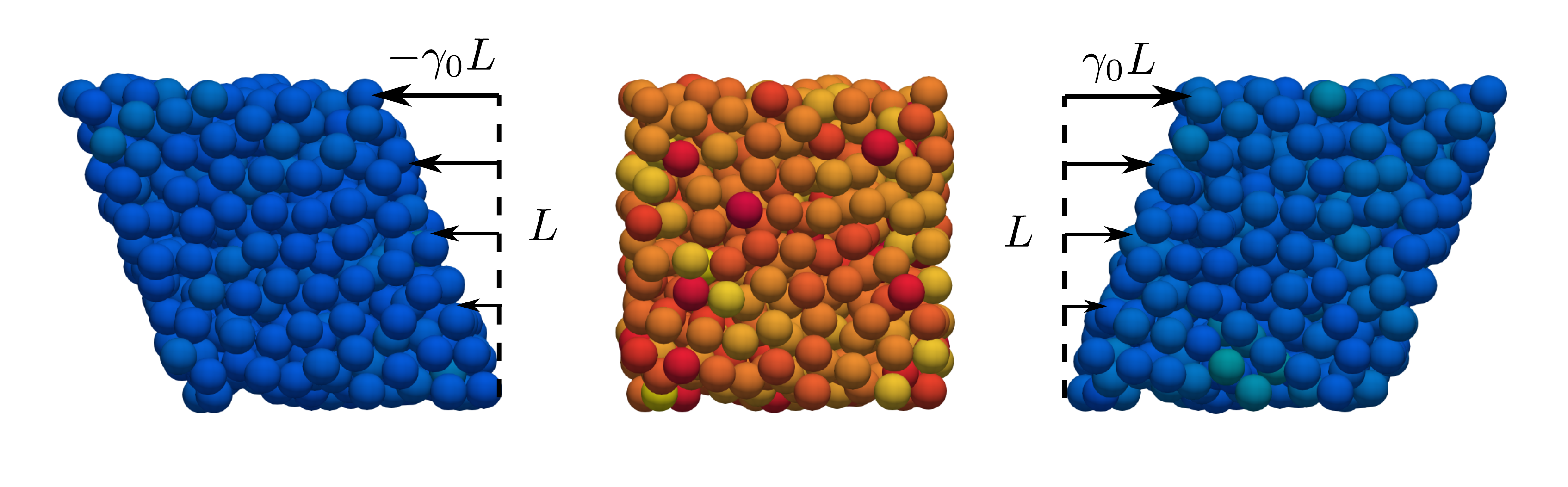}
	\caption{\re{Periodically sheared suspensions.
	             The schematic shows 500 particles in a cubic box of length $L$ at volume fraction $\phi=50\%$.
	             The particles are projected onto the shear plane,
	             where the color indicates the instataneous shear stress (reddish colors for higher stresses).
	             Here, the stresses are close to zero at maximum strains because the shear rate therein is zero.}}
	\label{fig:snaps}
\end{figure}

In our simulations we consider monodisperse suspensions made of 500 spheres 
(except in one case where 4096 particles are considered)
of radius $a$ at three volume fractions $\phi=30\%$, 40\%, and 50\%; \re{see Fig.~\ref{fig:snaps}}.
The computational domain is a cubic box with Lees-Edwards boundary conditions,
where the particles are driven by oscillatory strain, $\gamma(t) = \gamma_0 \sin(\omega t)$, 
under various driving amplitudes $\gamma_0$ and frequencies $\omega$.
This naturally defines a force scale, $F_\text{stk} \equiv 6\pi \eta_0 \gamma_0\omega a^2$,
being the Stokes drag of an isolated particle in a fluid with dynamic viscosity $\eta_0$.
Since particles in liquids are usually charged (due to surface acids or bases, adsorption of free ions, etc.),
we impose a pairwise repulsive force between nearby particles,
viz.~${\bm F}^\text{El} = {\bm F}_0 \exp(-\kappa h)$,
where ${\bm F}_0$ is the maximal repulsion,
$\kappa$ the inverse Debye length,
and $h$ the surface gap between two particles \cite{mewis_wagner_book, Israelachvili_book}.
$F_0$ can be equated with $F_\text{stk}$ to obtain a reference time scale to compare with that of the flow;
cf.~Ref.~\cite{ge2021irreversibility}.
Throughout this work we fix $F_0=10 F_\text{stk}$ and $\kappa^{-1}=0.01a$ to model a strong, short-range repulsion.
Effectively, this also keeps the non-dimensional shear rate (proportional to $F_\text{stk}/F_0$) 
constant under different $\gamma_0$ and $\omega$.
\re{The model produces quasi-Newtonian rheology with weak shear thickening under steady shear
(the suspension viscosity increases by less than a factor of 1.2 when the shear rate increases by 100 times).}
The source code of the FSD algorithm was originally published in the Supplementary Materials of Ref.~\cite{fiore2019fast}. 
The updated version we use is also available at \url{https://github.com/GeZhouyang/FSD}.

\section{Results}

In what follows, we discuss the rheology, collective dynamics, and underlying microstructural evolution of periodically sheared suspensions.
All simulations at the same volume fraction are started from the same random initial condition,
and the results are averaged over all particles (since they are identical)
and the last few cycles (if the  steady or quasi-steady state statistics are reported).
We have verified that the steady state results do not depend on the initial condition appreciably,
except when the suspension is driven by very small strain amplitudes in which case the microstructure does not evolve.
In general, we expect the insights gained from our analysis to be valid regardless of the model details.

\subsection{Rheology}

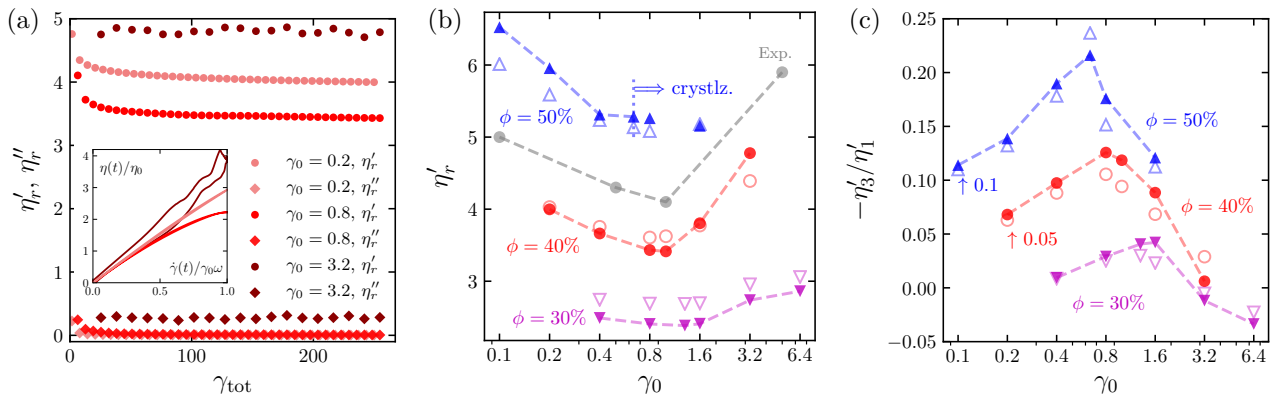
\begin{figure*}
	\centering
	\includegraphics[height=5.5cm]{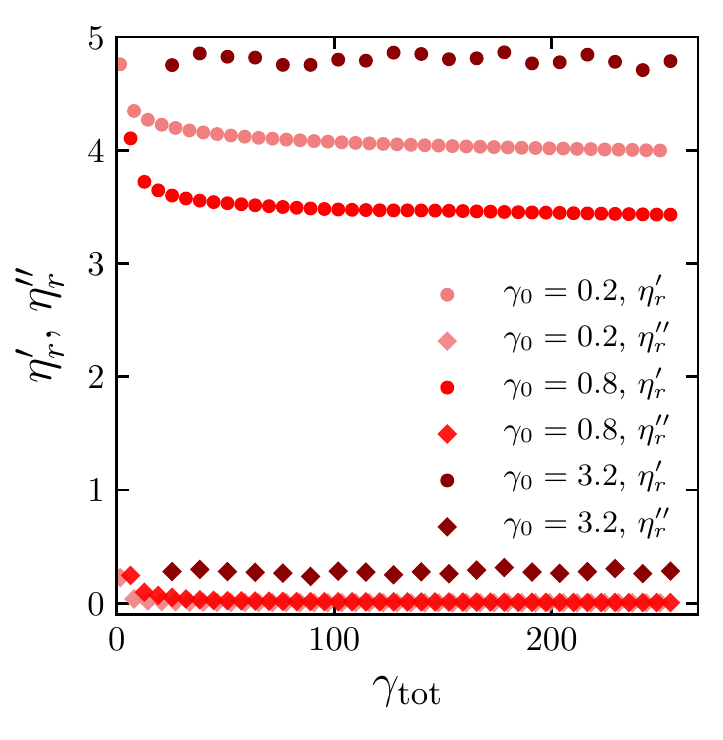}
	\includegraphics[height=5.5cm]{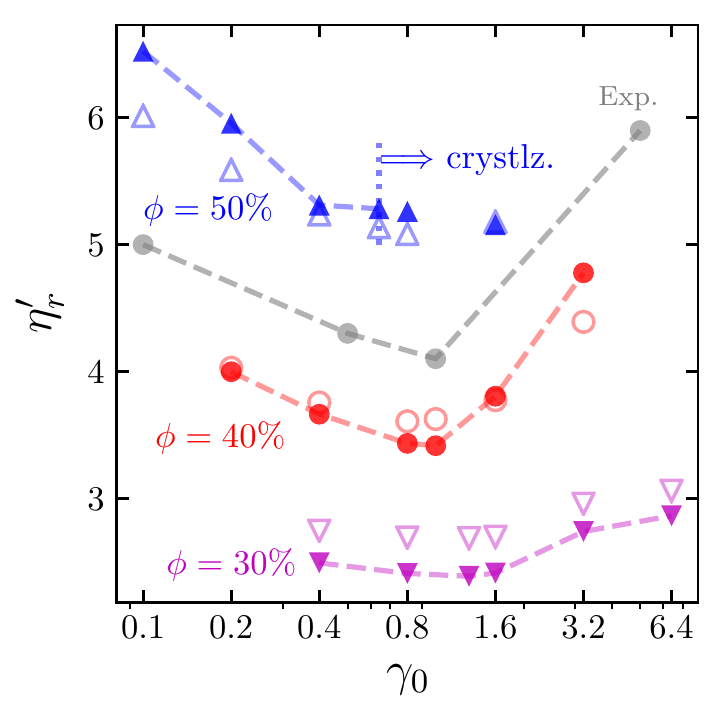}
	\includegraphics[height=5.5cm]{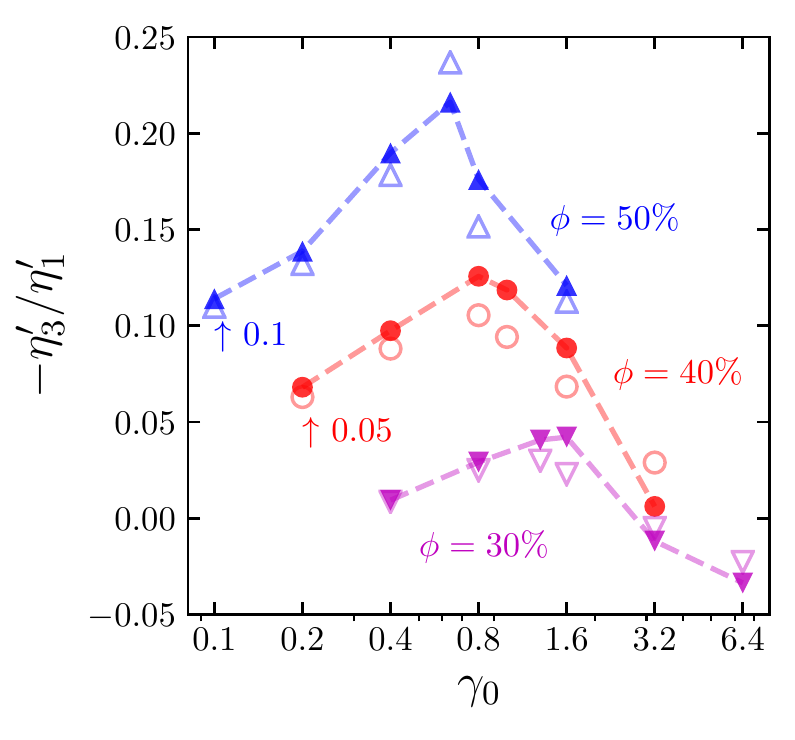}
	\setlength{\unitlength}{1cm}
	\begin{picture}(0,0)
		\put(-17.2,5.1){(a)} \put(-11.6,5.1){(b)} \put(-6,5.1){(c)}
		\put(-16.3,1.4){\includegraphics[height=2.2cm]{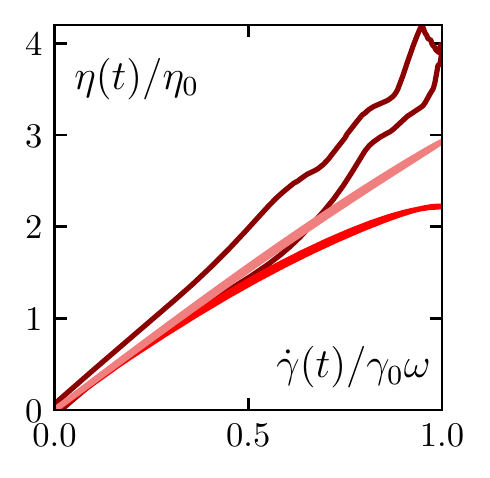}}
	\end{picture}
	\caption{Viscosity of periodically sheared suspensions.
	             (a) Evolution of the relative viscosity $\eta_r^\prime$ and relative elastic response $\eta_r^{\prime\prime}$ at $\phi=40\%$.
	             \re{Inset shows the corresponding total viscous responses during the last half cycle; see Eq.~\eqref{eq:stress-time-approx}.}
	             (b) Steady-state $\eta_r^\prime$ under various driving amplitudes and volume fractions.
	             \re{Gray circles are the experimental results at $\phi=40\%$ from Ref.~\cite{Bricker_Butler2006}.}
	             (c) Magnitude of the intracycle shear thinning quantified by $-\eta^\prime_3/\eta^\prime_1$ 
	             (results at $\phi=40\%$ and 50\% are shifted up by 0.05 and 0.1, respectively, for clarity).
	             In (b,c), filled symbols are with far-field hydrodynamic interactions, empty symbols without.}
	\label{fig:rheol}
\end{figure*}

\subsubsection{Viscosity}

One may decompose the shear stress $\sigma_{xy}(t)$ within an oscillation cycle in Fourier series,
\begin{equation} \label{eq:stress-time}
  \begin{aligned}
    \sigma_{xy}(t) = \gamma_0 \sum_{n=1}^{\infty} 
    \big(G_n^{\prime} \sin(n\omega t) + \omega\eta_n^\prime \cos(n\omega t)  \big), 
  \end{aligned}
\end{equation}
where $G_n^{\prime}$ and $\eta_n^\prime$ denote the $n$th elastic and viscous coefficients, respectively.
These coefficients vary from cycle to cycle;
but at steady state (when $\sigma_{xy}(t)$ becomes fully periodic), 
the leading-order even modes are always much smaller than their preceding odd modes, 
$G_{n-1}^{\prime}$ and $\eta_{n-1}^\prime$,
reflecting the fore-aft symmetry of the flow-induced microstructure \cite{gadala1980shear}.
Further inspection of the power spectrum shows that $\eta_n^\prime \gg G_n^{\prime}/\omega$ for all $n$,
as well as rapid decays of both $G_n^{\prime}$ and $\eta_n^\prime$ as $n$ increases 
(e.g.~$\eta_5^\prime/\eta_1^\prime \sim \mathcal{O}(10^{-2})$ typically).
These observations allow us to dramatically simplify Eq.~\eqref{eq:stress-time} 
and approximate the instantaneous shear viscosity as
\begin{equation} \label{eq:stress-time-approx}
  \begin{aligned}
    \eta(t) \approx  \eta_1^\prime \cos(\omega t) + \eta_3^\prime \cos(3\omega t), 
  \end{aligned}
\end{equation}
where $\eta \equiv \sigma_{xy} /\gamma_0\omega$.
In Eq.~\eqref{eq:stress-time-approx}, $\eta_1^\prime$ and $\eta_3^\prime$ correspond to 
the \emph{average} viscosity and the leading order \emph{intracycle} viscosity variation, respectively.
While $\eta_1^\prime$ is strictly positive for suspensions of passive particles,
$\eta_3^\prime$ can be either positive or negative.
Under the oscillatory strain $\gamma(t) = \gamma_0 \sin(\omega t)$, 
thus shear rate $\dot{\gamma}(t) = \gamma_0 \omega \cos(\omega t)$,
it can be shown that a negative (or positive) $\eta_3^\prime$ 
indicates shear thinning (or thickening) within a cycle \cite{ewoldt2008new}.
Note that this is not because $\eta_3^\prime$ is the next Fourier coefficient in Eq.~\eqref{eq:stress-time-approx},
but rather it is deduced from an orthogonal stress decomposition \cite{cho2005geometrical, ewoldt2008new}.

In the following, we rescale $\eta_1^\prime$ by the underlying fluid viscosity $\eta_0$ 
to obtain the conventional relative viscosity 
$\eta_r^\prime \equiv \eta_1^\prime/\eta_0$;
$\eta_3^\prime$ is rescaled by $-\eta_1^\prime$ to highlight the effect of shear thinning;
lastly, we define the relative elastic response (same unit as $\eta_r^\prime$) as 
$\eta_r^{\prime\prime} \equiv G_1^\prime/\eta_0\omega$. 

Fig.~\ref{fig:rheol}(a) shows the evolution of $\eta_r^\prime$ and $\eta_r^{\prime\prime}$,
\re{as well as the instantaneous viscous response to the imposed shear rate (obtaining the so-called Lissajous curves),}
under three representative strain amplitudes $\gamma_0=0.2$, 0.8, and 3.2 at $\phi=40\%$.
Here, the duration is measured in total strain units $\gamma_\text{tot}$ at the increment of full cycles;
i.e.~$\gamma_\text{tot}=4\gamma_0 n_\text{cyc}$, 
where $4\gamma_0$ is the accumulated strain over one cycle, and $n_\text{cyc}$ is the number of cycles.
\re{The Lissajous curves are plotted only during a half cycle, 
since both $\eta(t)$ and $\dot\gamma(t)$ are anti-symmetric about $t|_{\dot\gamma=0}$ at steady state.}
These plots reveal a few interesting features of the rheology under cyclic shear.
First, it confirms that the stress response is predominantly viscous, 
as $\eta_r^\prime \gg \eta_r^{\prime\prime}$ for all $\gamma_0$.
Second, it shows that the times to reach steady state are long;
there are continuous, albeit small, changes in both $\eta_r^\prime$ and $\eta_r^{\prime\prime}$ 
even after $\gamma_\text{tot} > 100$ for $\gamma_0 <3.2$.
\re{Third, the instantaneous viscous response within the last cycle shows qualitative differences under the three strain amplitudes:
the response is linear at $\gamma_0=0.2$, 
exhibits shear thinning at $\gamma_0=0.8$,
and becomes quasi-linear and hysteretic at $\gamma_0=3.2$.
The slow convergence to steady state and the three stress response patterns
are consistent with previous experiments and simulations 
\cite{gadala1980shear, Bricker_Butler2006, Martone_experiment, Ness_OS_2017SM},
suggesting that there might be a common rheological signature of periodically sheared suspensions.}
Furthermore, because stress is uniquely determined from microstructure in non-colloidal suspensions,
this implies glassy dynamics of suspensions under periodic strain,
which are indeed generally seen in disordered soft materials \cite{sollich1997rheology}.

To better examine the rheology, we show the steady-state $\eta_r^\prime$ and $-\eta_3^\prime/\eta_1^\prime$
under various strain amplitudes and volume fractions in Fig.~\ref{fig:rheol}(b,c).
Here, apart from the different behavior of $\eta_r^\prime$ in the densest packing 
where the suspension tends to crystallize under shear
\footnote{Shear-induced crystallization has been observed in monodisperse systems at high packing fractions,
cf.~Refs.~\cite{sierou2002rheology, duff2007shear, ghosh2022coupled}.
Numerically, this makes the steady state difficult to obtain as the structural changes are slow. 
The results for $\gamma_0 > 0.6$ at $\phi=50\%$ are thus taken at $\gamma_\text{tot}=256$ rather than the true steady states.},
both viscous coefficients are non-monotonic in $\gamma_0$:
at each $\phi$, the relative viscosity tends to a minimal value near an intermediate amplitude,
where the effect of intracycle shear thinning is roughly at a maximum \re{(cf.~Fig.~\ref{fig:rheol}(a), inset)}.
$\gamma_0$-dependent viscosity has been previously observed in oscillatory shear experiments
\cite{breedveld2001shear, Bricker_Butler2006, Lin_Phan-Thien_Khoo_2013, Ness_OS_2017SM, Martone_experiment}.
\re{For example, Bricker and Butler performed extensive experiments of particle suspensions at $\phi=40\%$,
using four different types of materials in two rheometer geometries, 
and found that the complex viscosity is smallest at $\gamma_0\approx1$ \cite{Bricker_Butler2006,Bricker_Butler2007}.
Our results are in qualitative agreement with theirs
(the precise value of $\eta_r^\prime$ depends not only on $\gamma_0$, 
but also on the physicochmeical interaction between the particles); see Fig.~\ref{fig:rheol}(b).}

The physical origin of the observed shear rheology lies in the structural changes caused by \emph{repeated} flow reversals.
Microscopically, shearing a suspension creates a non-uniform structure characterized by anisotropic pair distributions,
which are correlated with higher particle stresses \cite{brady1997microstructure, sierou2002rheology, morris2009review}.
The deviation of this shear-induced microstructure from a given initial configuration, 
as well as the extent to which it can be reversed within a cycle, 
depends on the applied strain and the straining \emph{history}.
\re{Crucially, history matters only transiently and when $\gamma_0$ is ``small".
The microstructure formed under large $\gamma_0$ is statistically invariant within each half cycle;
that is, it always reaches the same steady state twice every period.
The threshold $\gamma_0$ above which history plays diminishing role in the rheology is related to the RIT
and will be quantified in Sec.~\ref{sec:dyn}.
As we shall see, it is the combined effects of shearing with a finite amplitude and the resulting microstructural evolution
that give rise to the non-monotonic viscosity variations on average and within a cycle.}

\subsubsection{Normal stress differences}

\re{Apart from a varying viscosity, suspensions can also develop anisotropic normal stresses under shear,
leading to finite normal stress differences (NSDs).
Rheologically, NSDs account for a number of non-Newtonian phenomena (most notably the Weissenberg effect)
and are a consequence of the flow-induced microstructure \cite{morris2009review}.
Below, we examine NSD responses of periodically sheared suspensions.}

\re{Similar to the shear stress, any component of the normal stresses, $\sigma_{\alpha\alpha}(t)$, 
may be decomposed within a cycle in Fourier series,
\begin{equation} \label{eq:nor_stress-time}
  \begin{aligned}
    \sigma_{\alpha\alpha}(t) = \sum_{n=0}^{\infty} \hat{\sigma}_n^{(\alpha)} \exp(i n\omega t) , 
    \quad \alpha = x \textrm{ or } y \textrm{ or } z,
  \end{aligned}
\end{equation}
where $\hat{\sigma}_n^{(\alpha)}$ denotes the $n$th complex Fourier coefficient of $\sigma_{\alpha\alpha}$.
The first and second NSD, defined respectively as 
$N_1\equiv \sigma_{xx}-\sigma_{yy}$ and $N_2\equiv \sigma_{yy}-\sigma_{zz}$,
are thus also decomposed 
\begin{subequations} 
  \begin{align}
     N_1(t) = \sum_{n=0}^{\infty} \big(\hat{\sigma}_n^{(x)}-\hat{\sigma}_n^{(y)} \big) \exp(i n\omega t), \label{eq:N1-time}\\   
     N_2(t) = \sum_{n=0}^{\infty} \big(\hat{\sigma}_n^{(y)}-\hat{\sigma}_n^{(z)} \big) \exp(i n\omega t). \label{eq:N2-time}
  \end{align}
\end{subequations}
Note that, NSDs can have nonzero averages over a cycle and are even functions of the applied strain at steady state.
This is in contrast to the shear stress, though the different behaviors are all ultimately due to the underlying time-reversal symmetry.}

\begin{figure}
	\centering
	\includegraphics[height=6.5cm]{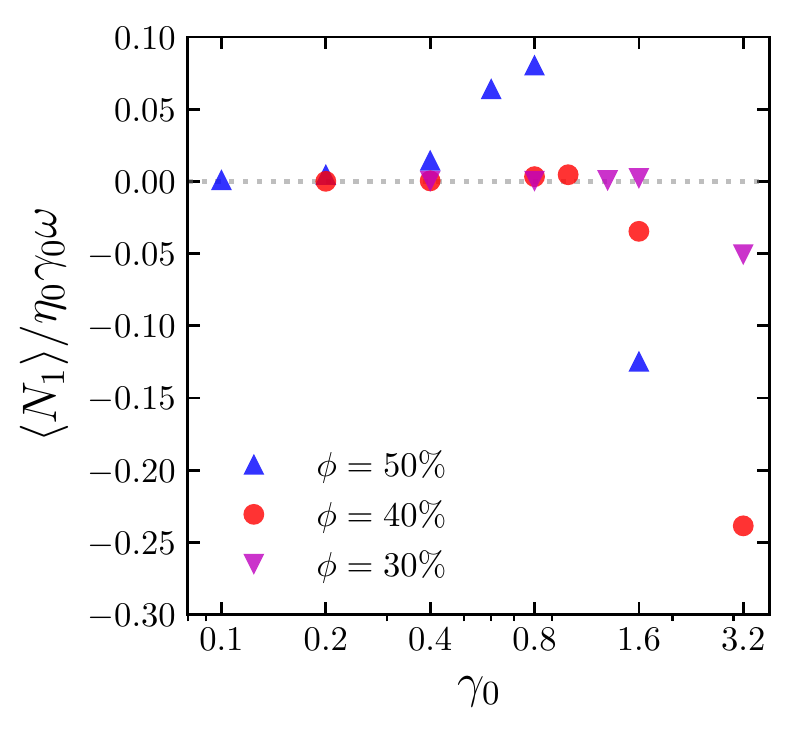}
	\setlength{\unitlength}{1cm}
	\begin{picture}(0,0)
		\put(-5.5,1.2){\includegraphics[height=3cm]{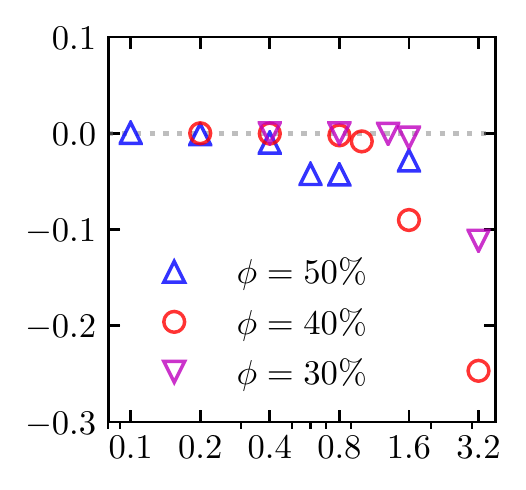}}
	\end{picture}
	\caption{\re{Average responses of the first normal stress difference of periodically sheared suspensions.
	              Inset shows the pure hydrodynamic contributions.}}
	\label{fig:N1}
\end{figure}

\re{Fig.~\ref{fig:N1} shows the average $N_1$ normalized by $\eta_0\gamma_0\omega$
under various $\gamma_0$ and $\phi$ at steady state.
The most salient feature is that $N_1$ remains negligible for small $\gamma_0$
until it becomes finite as $\gamma_0$ further increases.
A negligible $N_1$ suggests that the pair distribution function characterizing suspension microstructure is nearly fore-aft symmetric,
as would be the case in the purely hydrodynamic limit \cite{morris2009review}.
This ceases to be the case when $\gamma_0$ is beyond a $\phi$-dependent threshold,
where $N_1$ first becomes positive (ever so slightly), then negative.
A positive $N_1$ can be due to either a higher tensile normal stress in the flow direction, 
or more collisions (which contribute to a negative normal stress) in the velocity gradient direction.
Our simulations suggest the latter, as the hydrodynamic contributions to $N_1$ are always negative (see Fig.~\ref{fig:N1}, inset).
Lastly, we note that a transition from positive to negative $N_1$ has been observed before 
and may have certain rheological implications \cite{Bricker_Butler2007}.
Whether this is a generic behavior or can be neglected remains to be explored.}

\begin{figure*}
	\centering
	\includegraphics[height=5.5cm]{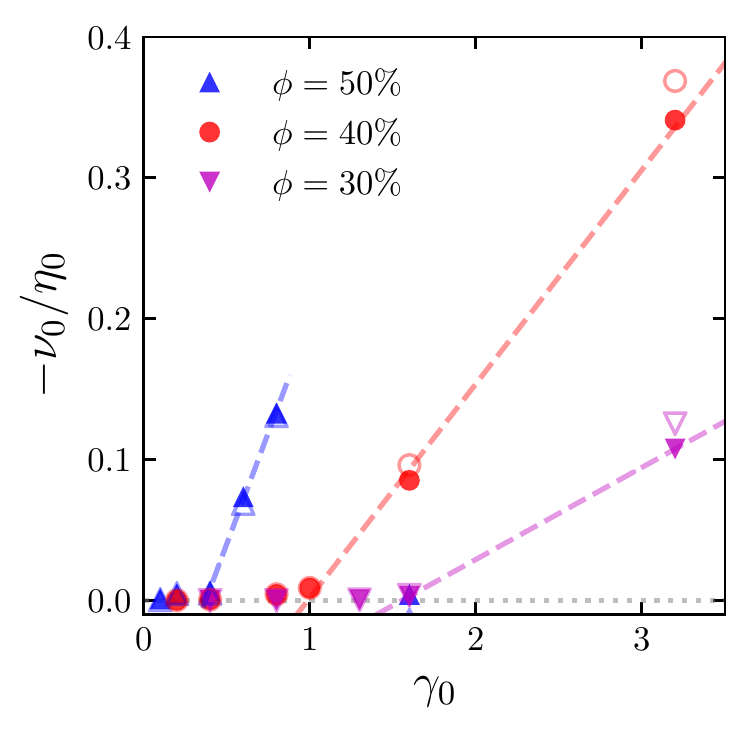}
	\includegraphics[height=5.5cm]{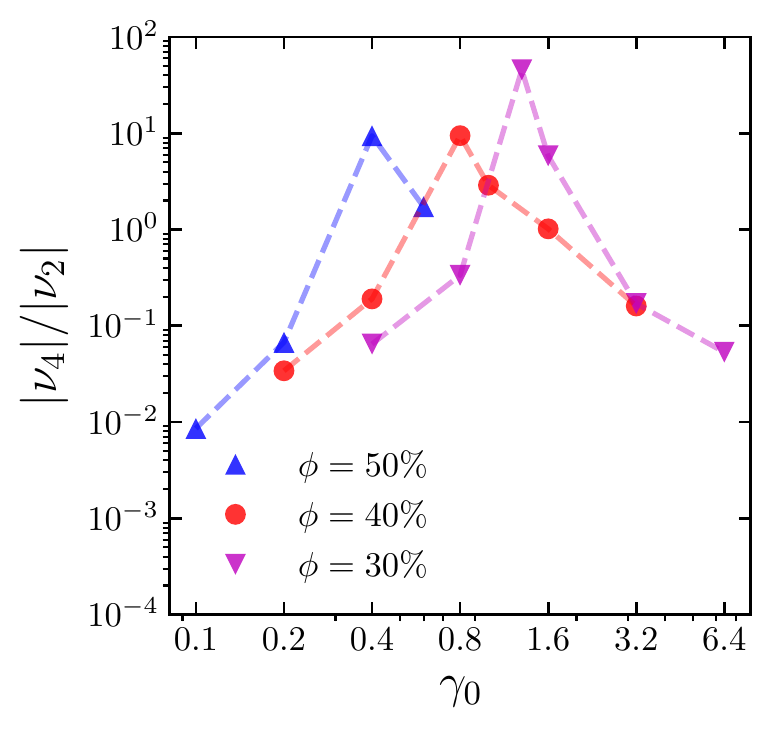}
	\includegraphics[height=5.5cm]{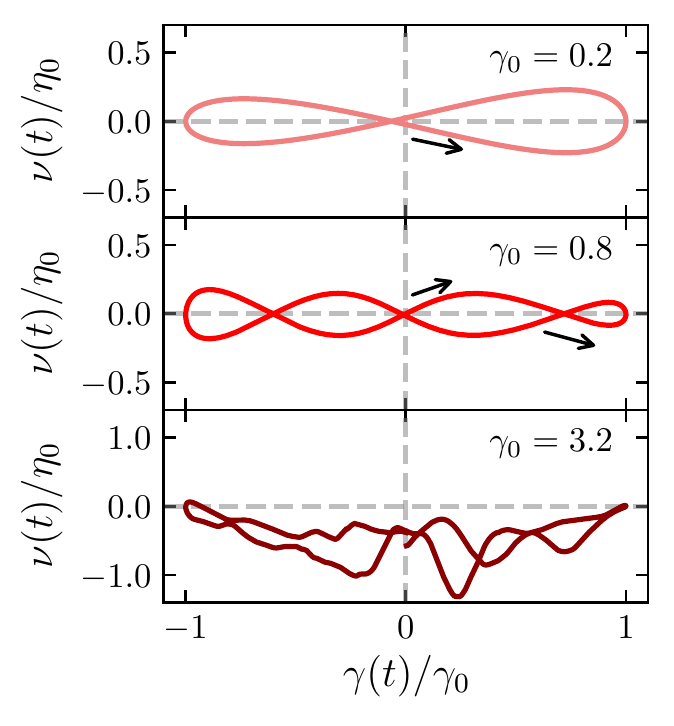}
	\setlength{\unitlength}{1cm}
	\begin{picture}(0,0)
		\put(-16.9,5.1){(a)} \put(-11.1,5.1){(b)} \put(-5.5,5.1){(c)}
		\put(-8,1){\includegraphics[height=1.9cm]{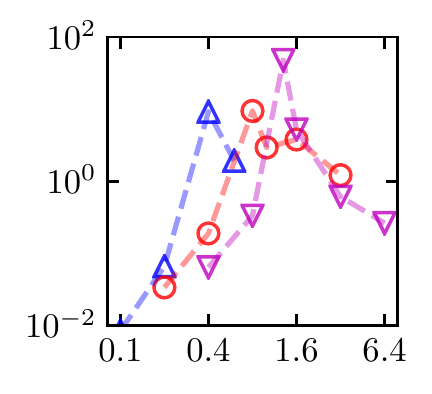}}
	\end{picture}
	\caption{\re{The second normal stress difference of periodically sheared suspensions. 
	             (a) Average $N_2$ per cycle, where the dashed lines are linear fits of the last few data points
	             (filled symbols are the entire stress, empty symbols are contributions from collisions and electrostatic repulsions).
	             (b) Ratio of the magnitudes of the fourth ($|\nu_4|$) and second ($|\nu_2|$) harmonic responses 
	             (inset shows the ratio due to hydrodynamic interactions). 
	             (c) Lissajous curves at $\phi=40\%$ for three representative $\gamma_0$ (arrows show the direction of the curves).}}
	\label{fig:N2}
\end{figure*}
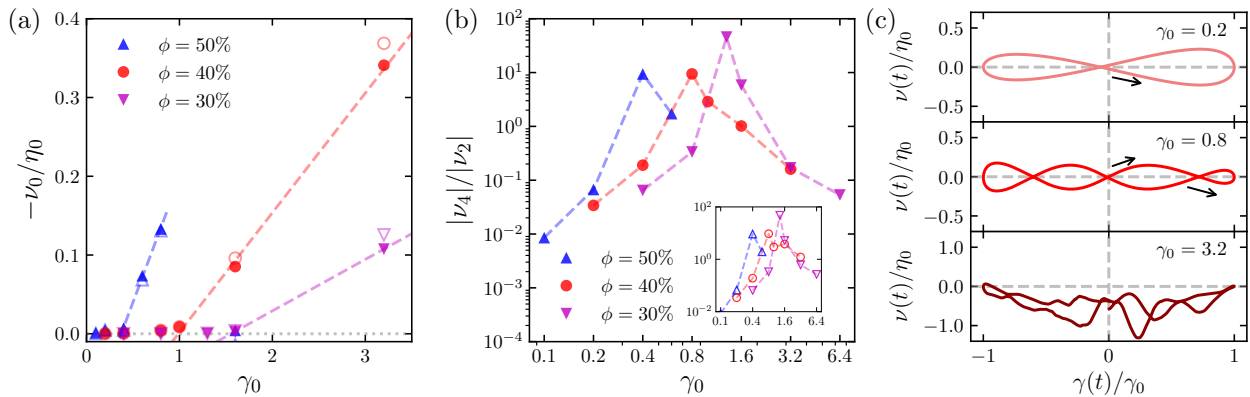

\re{We now turn to $N_2$, 
which is consistently reported negative and dominates over $N_1$ (i.e.~$|N_2|>|N_1|$) for suspensions \cite{guazzelli_pouliquen_2018},
but has received somewhat less attention \cite{maklad2021review}.
Dividing $N_2$ by $\gamma_0\omega$ and keeping three leading-order even modes, 
its instantaneous response is approximated as
\begin{equation} \label{eq:nor_stress-time-approx}
  \begin{aligned}
    \nu(t) \approx \nu_0 + \nu_2\exp(2i\omega t) + \nu_4\exp(4i\omega t),
  \end{aligned}
\end{equation}
where $\nu \equiv N_2 /\gamma_0\omega$.
Note that, the coefficients $\nu_0$, $\nu_2$, and $\nu_4$ are not necessarily in decaying magnitude,
but the higher order terms are always decaying (i.e.~$|\nu|_{n+2} \ll |\nu|_n$, $n=4,6,8,\dots$) upon inspection of the power spectrum.
In the following, we rescale $\nu_0$ by $-\eta_0$ to examine the average $N_2$ per cycle,
and compare the magnitudes of the second ($|\nu_2|$) and fourth ($|\nu_4|$) harmonic responses to analyze the nonlinear response.}

\re{Fig.~\ref{fig:N2}(a) shows the average $N_2$ for all $\phi$ and $\gamma_0$ at steady state.
There are clearly two regimes: 
under small $\gamma_0$, $N_2 \approx 0$ as in the case of $N_1$,
indicating an overall isotropic microstructure in all principal directions;
under large $\gamma_0$, $N_2 < 0$ as particles collide more in the shear plane than in the flow-vorticity plane,
consistent with typical observations in simple shear flow \cite{guazzelli_pouliquen_2018} (notice the hollow markers in Fig.~\ref{fig:N2}a).
The onset of a finite $N_2$ occurs at a well-marked $\gamma_0$,
which reduces with $\phi$ and, importantly,
is within the range of the corresponding $\gamma_0$ where the minimum $\eta_r^\prime$ (prior to crystallization, if any) 
and the maximum $-\eta_3^\prime/\eta_1^\prime$ are also obtained, cf.~Fig.~\ref{fig:rheol}(b,c).
As will be shown in Sec.~\ref{sec:dyn},
this is not a coincidence but rather an indication of RIT.}

\re{The most unexpected behavior of $N_2$ is its nonlinear response near the threshold strain amplitude.
As shown in Fig.~\ref{fig:N2}(b), this nonlinear effect is so strong that $|\nu_4|/|\nu_2|$ can increase by three orders-of-magnitude.
Therefore, the responding frequency of $N_2$ under an imposed frequency $\omega$ will double for a range of $\gamma_0$ (i.e.~$2\omega \to 4\omega$).}

\re{To understand this behavior, we examine the Lissajous curves of the normalized instantaneous second normal stress difference ($\nu(t)/\eta_0$) 
vs.~strain ($\gamma(t)/\gamma_0$) for three representative $\gamma_0$ at $\phi=40\%$ in Fig.~\ref{fig:N2}(c).
At $\gamma_0=0.2$, $\nu(t)/\eta_0$ exhibits the expected periodic response with a $2\omega$ frequency:
it is anti-symmetric about both the zero strain (due to reversibility) and the zero shear rate (due to even symmetry). 
At $\gamma_0=3.2$, $\nu(t)/\eta_0$ is no longer zero at the zero strain 
(due to persistent particle collisions, consistent with its non-zero average), 
but still responds to the driving strain at $2\omega$.
Interestingly, at $\gamma_0=0.8$ where the response frequency doubles,
$\nu(t)/\eta_0$ is initially positive as $\gamma(t)/\gamma_0$ increases from 0,
but only up to a fraction of the strain amplitude, 
then it becomes negative as $\gamma(t)/\gamma_0$ further increases; 
see arrows in Fig.~\ref{fig:N2}(c).
A positive $N_2$ can only be due to higher tension in the direction aligning with the velocity gradient than the vorticity,
since collisions occur more frequently in the shear plane than the others, as mentioned earlier.
Indeed, the stress contributions from hydrodynamic interactions alone exhibit the same nonlinear response (see Fig.~\ref{fig:N2}(b) inset).
This is in stark contrast to the average response, which is dominated by the interparticle repulsion and collisions.
}

\re{Finally, we note that all aforementioned rheological responses, namely, 
the reduced relative viscosity,
the enhanced intracycle shear thinning, 
the onset of a finite second normal stress difference and 
its frequency doubling within a cycle,
\emph{all} occur within a narrow range of $\gamma_0$ that depends on $\phi$ only.
We will show that the $\phi$-dependent $\gamma_0$ coincides with the critical strain amplitude $\gamma_c$ of RIT next.}

\subsection{Dynamics}
\label{sec:dyn}

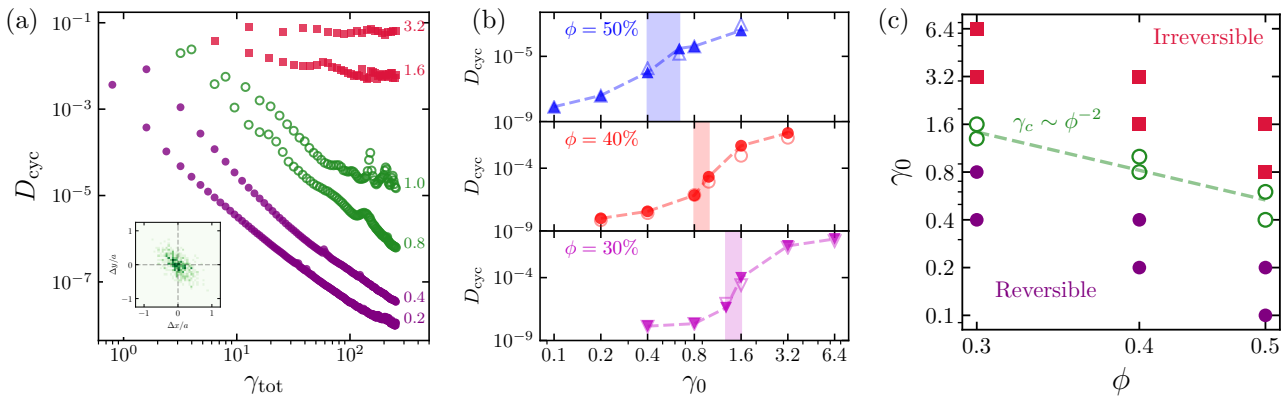
\begin{figure*}
	\centering
	\includegraphics[height=5.5cm]{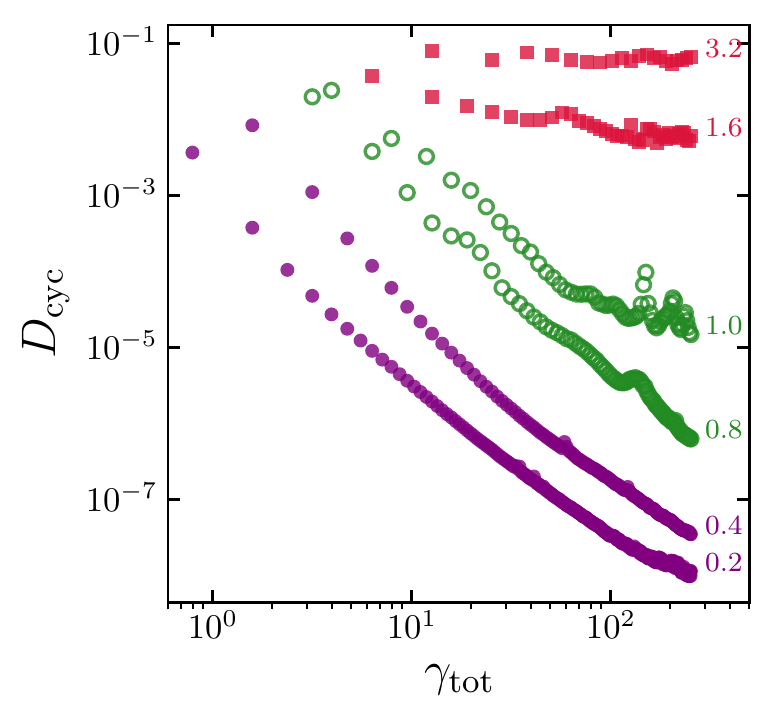}
	\includegraphics[height=5.5cm]{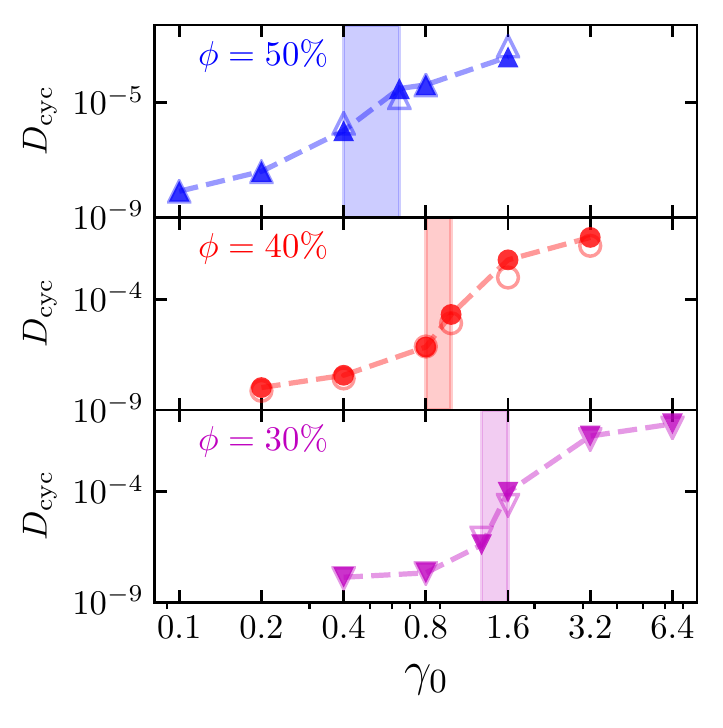}
	\includegraphics[height=5.5cm]{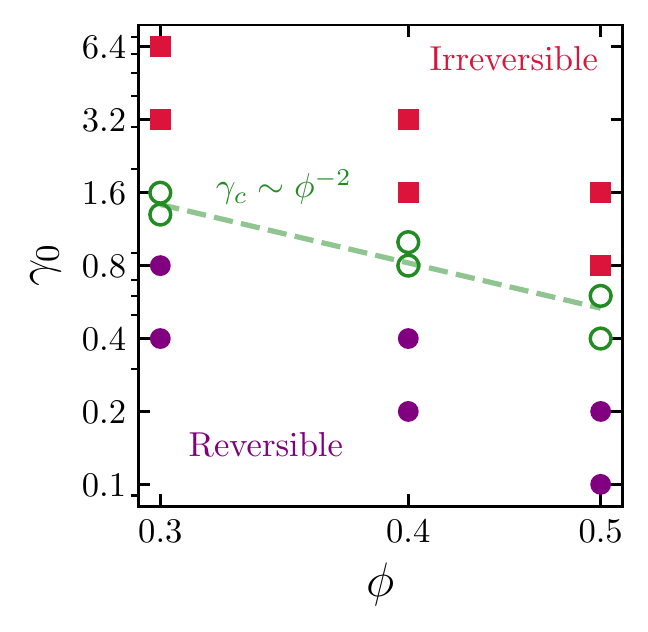}
	\setlength{\unitlength}{1cm}
	\begin{picture}(0,0)
		\put(-17.4,5.1){(a)} \put(-11.2,5.1){(b)} \put(-5.8,5.1){(c)}
		\put(-16.1,1){\includegraphics[height=1.6cm]{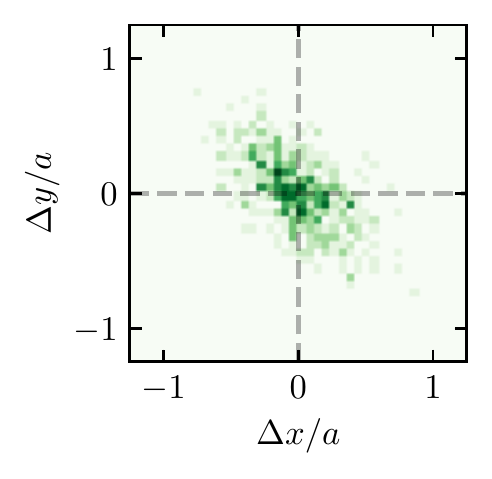}}
	\end{picture}
	\caption{Dynamics of periodically sheared suspensions.
	             (a) Evolution of the per-cycle diffusivity $D_\text{cyc}$ in total strain $\gamma_\text{tot}$ 
	             under various strain amplitudes $\gamma_0$ (labelled aside) at $\phi=40\%$.
	             Inset shows the particle dispersion under $\gamma_0=0.8$ after half a cycle.
	             (b) Final $D_\text{cyc}$ evaluated at $\gamma_\text{tot}=256$ under various $\gamma_0$ and $\phi$
	             (filled symbols are with far-field hydrodynamic interactions, empty symbols without).
	             The shaded regions correspond to the range of $\gamma_0$ 
	             \re{estimated from the rheological signatures 
	             according to Figs.~\ref{fig:rheol}(b,c) and \ref{fig:N2}(a,b) (see text).}
	             (c) Phase diagram of the reversible-irreversible transition.
	             The empty circles are estimated bounds (shades in b) of the critical amplitude $\gamma_c$,
                     whereas the dashed line corresponds to $\gamma_c = 0.14\phi^{-1.93}$ \cite{Pine_Nature_2005}.}
	\label{fig:dyn}
\end{figure*}

The $\gamma_0$-dependent transient and steady-state rheology implies irreversible particle dynamics.
If the particle trajectories are completely reversible from the beginning, 
we should not expect to observe any stress evolution from cycle to cycle,
\re{and both $\eta_r^\prime$ and $-\eta_3^\prime/\eta_1^\prime$ are likely to change monotonically with $\gamma_0$
(assuming the starting configuration is random).}
Irreversibility, however, may last only \emph{transiently} 
as particles can self-organize into reversible absorbing states 
according to a so-called random organization (RO) model \cite{Corte_NatPhys_2008, Wilken2021random}.
In that model, the only physical assumptions are 
(I) particles do not overlap and 
(II) there is some source of microscopic irreversibility (or noise).
Despite the simplicity, the model applies to a wide range of systems subject to cyclic shear,
and is believed to be the underlying reason for the observed reversible-irreversible transition.

To check irreversibility, we sample the particle statistics directly from their trajectories in the simulations.
Fig.~\ref{fig:dyn}(a) shows the transient self-diffusivity under various strain amplitudes $\gamma_0$ at volume fraction $\phi=40\%$,
where the diffusivity $D_\text{cyc}$ is computed \emph{per cycle} according to
\begin{equation} \label{eq:Diff}
     \langle (\Delta_i/a)^2 \rangle = 2 D_\text{cyc} \gamma_\text{cyc},
\end{equation}
where $\Delta_i$ is the displacement of particle $i$ over a cycle, 
$\langle \cdot \rangle$ averages over all particles, 
and $\gamma_\text{cyc}=4\gamma_0$.
Clearly, the initial dynamics are \re{irreversible} under all amplitudes.
As a further illustration, 
the inset in Fig.~\ref{fig:dyn}(a) shows the distribution of particle dispersions after the first \emph{half} cycle under $\gamma_0=0.8$.
Here, the particles not only fail to return to their original positions,
but are also displaced anisotropically due to shear.
After another half cycle, the preferential direction of the dispersion will reverse (i.e.~$\Delta \to -\Delta$).
Repeated oscillations eventually leads to either decaying or persistent diffusion,
depending on $\gamma_0$,
as predicted by the RO model.

The final per-cycle diffusivity is plotted as a function of strain amplitudes for all volume fractions in Fig.~\ref{fig:dyn}(b),
where $D_\text{cyc}$ is evaluated at $\gamma_\text{tot}=256$.
\re{The common pattern of these curves is a transition from reversible to irreversible dynamics as $\gamma_0$ increases.
For each $\phi$, we highlight the range of $\gamma_0$ 
within which 
the suspension exhibits the aforementioned rheological signatures illustrated in Figs.~\ref{fig:rheol} and \ref{fig:N2}.}
This is done by intersecting the range of $\gamma_0$ estimated individually.
For example, at $\phi=40\%$, we estimate that the minimum $\eta_r^\prime$ is obtained under $\gamma_0 \in (0.8,1.6)$,
whereas the maximum $-\eta_3^\prime/\eta_1^\prime$ is obtained under $\gamma_0 \in (0.4,1.0)$;
thus, the overlap $\gamma_0 \in (0.8,1.0)$ provides a bound for the critical amplitude $\gamma_c$.
\re{The onset of $-\nu_0/\eta_0$ and the peak of $|\nu_4|/|\nu_2|$ also fall in the same range.} 
Despite the sparsity of data, these bounds roughly correspond to the region where the dynamics are the most susceptible to changes in $\gamma_0$,
i.e.~where $\partial D_\text{cyc}/\partial \gamma_0$ is the steepest.
Therefore, our results suggest that rheological measures can also be used to infer the dynamical phase transition of suspensions under cyclic shear,
as plotted in Fig.~\ref{fig:dyn}(c).

The critical amplitude $\gamma_c$ for periodically sheared suspensions considered herein depends only on the volume fraction.
Previous experiments report $\gamma_c = 0.14\phi^{-1.93}$ by extrapolating the effective particle diffusivities 
\re{(based on particle mean square displacements over all cycles, rather than per cycle)}
from $\gamma_0>\gamma_c$ \cite{Pine_Nature_2005}.
In our simulations it is difficult to track the particle displacements under large $\gamma_0$ over long time (e.g.~$\gamma_\text{tot}>256$) 
due to the finite size of the computational domain.
However, our rheologically estimated $\gamma_c$ agrees rather well with the experimental fit,
as seen in Fig.~\ref{fig:dyn}(c);
hence, it validates the alternative approach.

The scaling of $\gamma_c$ with respect to $\phi$ can be explained qualitatively by the phenomenon of shear-induced diffusion.
Earlier experiments on non-colloidal suspensions in simple shear flow found that 
the self-diffusivity $D_s$ scales with the particle size $a$, the imposed shear rate $\dot{\gamma}$, and the volume fraction $\phi$ as
$D_s \sim \phi^2\dot{\gamma}a^2$ \cite{leighton_acrivos_1987}.
Here, the $\phi^2$ dependence is due to many-body interactions,
since the probability of one particle interacting with two others is proportional to $\phi^2$ in a statistically uniform suspension
\re{(pair interactions would lead to a scaling of $D_s \sim \phi\dot{\gamma}a^2$,
but they are insufficient to cause persistent diffusion when the non-hydrodynamic force is weak \cite{drazer_koplik_khusid_acrivos_2002}).}
Replacing $\dot{\gamma}$ with $\gamma_0\omega$ and rescaling $D_s$ by $a^2\omega$, 
we have $\tilde{D}_s \equiv D_s/(a^2\omega) \sim \phi^2\gamma_0$;
thus the scaling of $\gamma_c \sim \phi^{-2}$ at a measurable onset diffusivity.
Counterintuitively, our simulations suggest that far-field hydrodynamic interactions are \emph{not} responsible for the many-body effect,
as the results are qualitatively the same with and without such interactions;
see Figs.~\ref{fig:rheol}(b,c) and \ref{fig:dyn}(b).
Effectively, local interactions such as the lubrication and electrostatic repulsion are sufficient 
to couple the particle motions to yield the observed rheology and collective dynamics in concentrated suspensions.
By contrast, far-field hydrodynamic interactions promote reversibility 
(evidently in the extreme case of two particles in shear flow \cite{batchelor1972hydrodynamic, da1996shear}),
thus acting to smear the dynamical phase transition \cite{metzger2010irreversibility}.

\re{Finally, we note that the consequence of RIT 
on the rheology is non-trivial.
When $\gamma_0 < \gamma_c$,
the per-cycle dynamics are not immediately quiescent, but rather decay in time,
thus changing the reference condition at the beginning of each cycle;
this is manifested in the intracycle rheological responses of both the viscosity and second normal stress difference.
When $\gamma_0 > \gamma_c$,
the particle dynamics depend less and less on the strain history as collisions become persistent, 
leading to anisotropic microstructures that are evident in the increased viscosity and finite second normal stress difference.
These two regimes have distinct rheological behaviors which, at the onset of transition, is highly nonlinear.
Below, we examine the suspension microstructure at $\gamma_0 \approx \gamma_c$ 
to elucidate the connection between the rheology and emergent dynamics.}

\subsection{Microstructure}
\label{sec:micro}

\begin{figure}
	\centering
	\includegraphics[width=4.25cm]{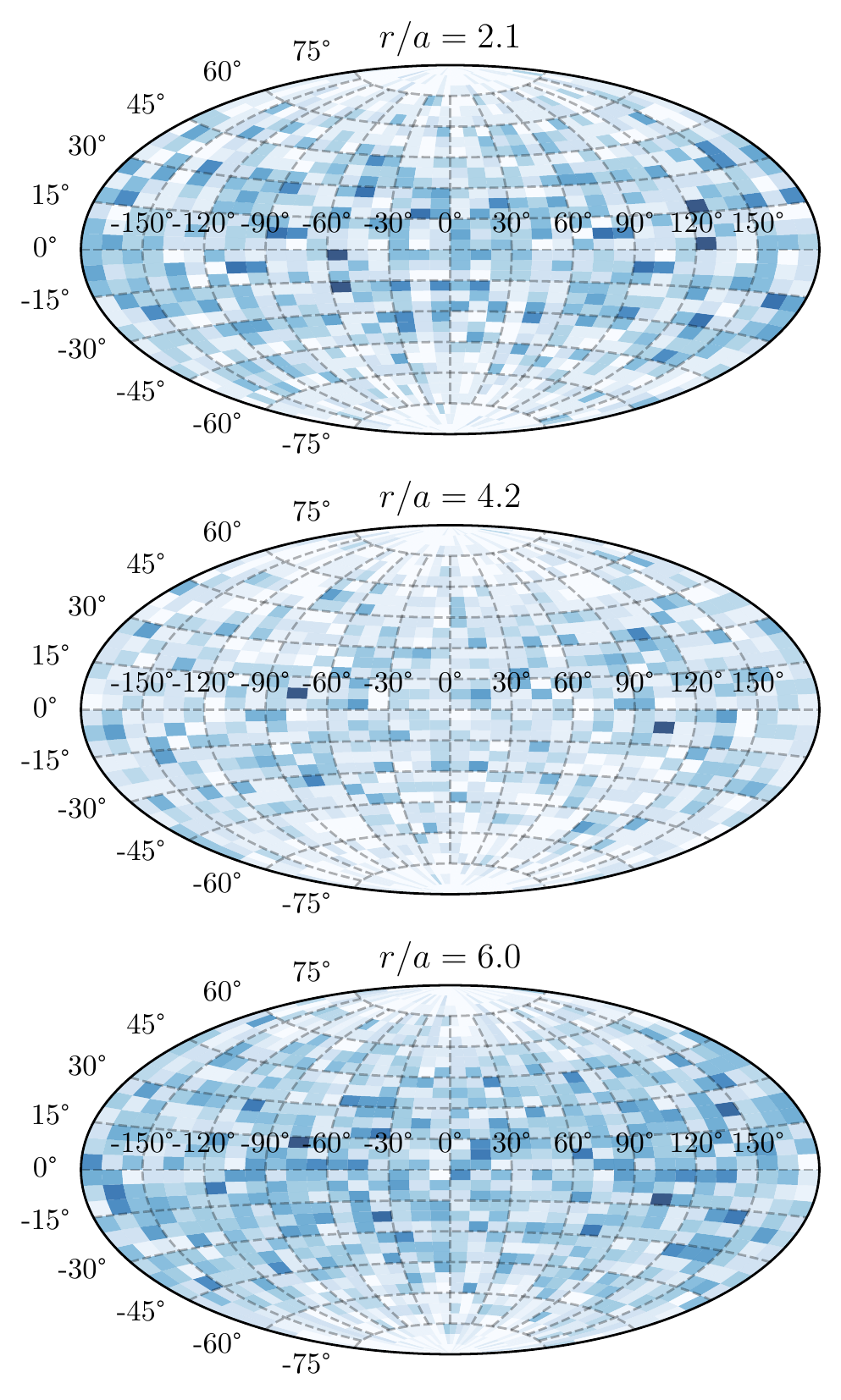}
	\includegraphics[width=4.25cm]{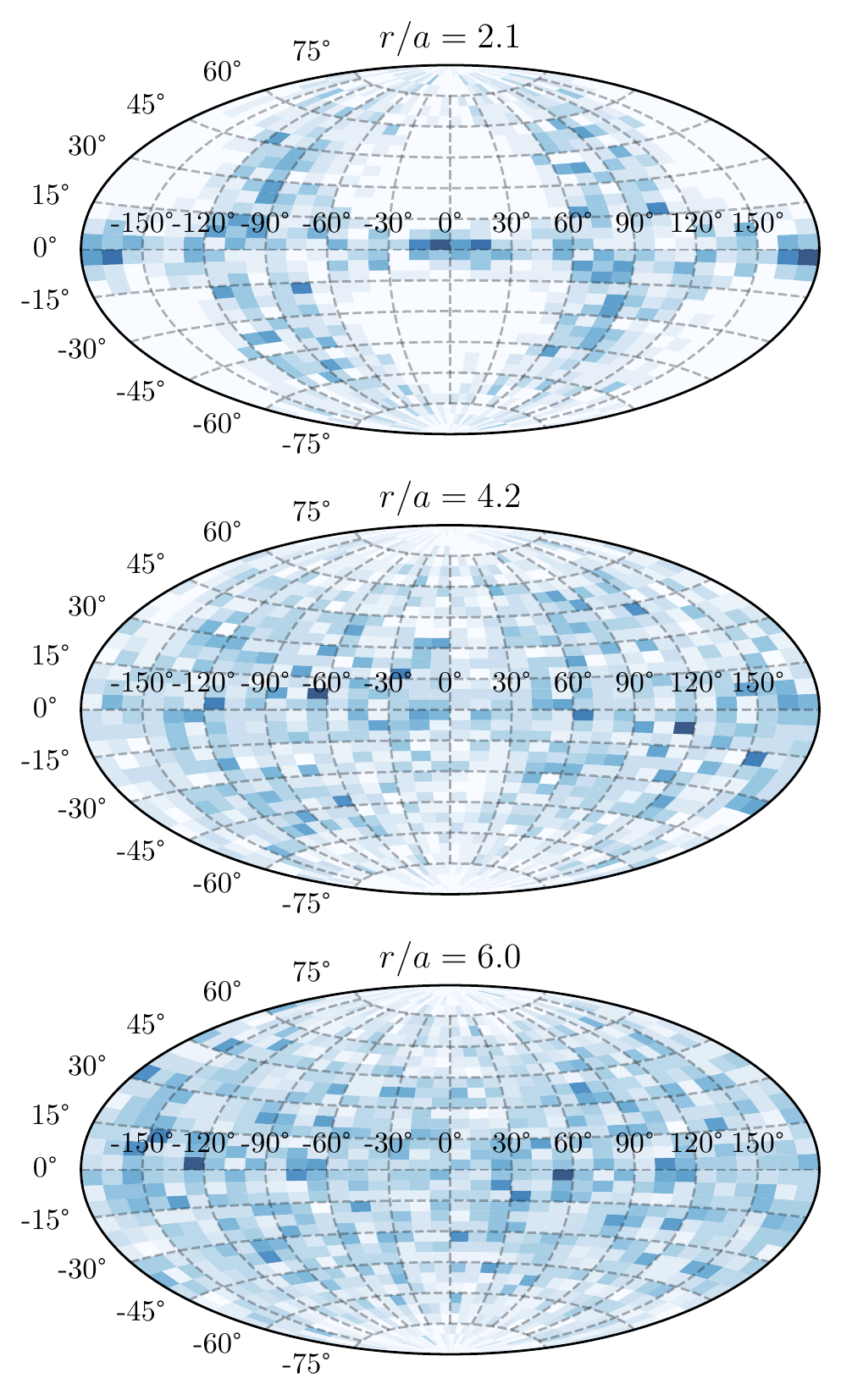}
	\caption{Particle angular distributions (Aitoff projection) under $\phi=40$\% and $\gamma_0=0.8$.
              	     (left) Initial distributions. (right) Final distributions.
                     In all cases, the angular distribution is averaged at distance $r \pm 0.1a$,
                     where darker colour indicates higher probability.
                     The maps are oriented such that the shear flow has the same velocity along latitudes.}
	\label{fig:map}
\end{figure}

The correspondence between rheology and dynamics can only be due to the underlying microstructure,
which may be characterized by the pair correlation function.
As an example, we show in Fig.~\ref{fig:map} the particle angular distributions under $\gamma_0=0.8$ and $\phi=40$\%,
at three representative distances corresponding to the first three peaks of the radial distribution function at steady state.
The initial random configuration shows the expected isotropic distribution at all distances.
However, the final distribution within a boundary layer of thickness $0.1a$ is strongly anisotropic,
with virtually no particles at latitudes above $15\degree$ or below $-15\degree$,
except for those near longitude $\pm 90\degree$.
Here, the equatorial ring corresponds to regions of the smallest velocity difference,
whereas the longitudinal ring is perpendicular to the flow;
particles in both of these regions experience less relative motion under shear than elsewhere \cite{da1996shear}.
Since particles tend to be displaced in the velocity gradient direction under shear (see the inset in Fig.~\ref{fig:dyn}a),
repeated oscillations lead to a local depletion of particles from regions of greater encounter probability.
The final effect is the formation of a highly correlated short-range structure, 
with seemingly little change in the bulk, as depicted in Fig.~\ref{fig:map}.
We observe this phenomenon in general for all cases;
however, it is always the most pronounced at $\gamma_0 \approx \gamma_c$.

To gain further insight into the bulk structure near the critical transition point, 
we analyse the nearest-neighbour graph (NNG) obtained from snapshots of the particle configuration.
By definition, a NNG is formed by connecting each particle to its nearest neighbour,
resulting in a partition of the particles into clusters (or minimum spanning trees as in graph theory) of various size and shape.
Specifically, we define the size as the number of particles in the cluster,
and characterize the shape from the moment-of-inertia tensor ${\bm I}_c$ of the (as if rigid) cluster;
that is, we define the shape factor as $\sqrt{I_{max}/I_{min}}$,
where $I_{max}$ and $I_{min}$ are the maximum and minimum eigenvalues of ${\bm I}_c$, respectively.
Under these definitions, the minimum size is 2 (i.e.~a doublet),
while the minimum shape factor is 1 (i.e.~an isotropic cluster).
NNG is suitable for clustering analysis due to its simplicity and analytical properties \cite{eppstein1997nearest};
here, we use it also because it is a unique and parameter-free characterization of the suspension network.

\begin{figure}
	\centering
	\includegraphics[height=8cm]{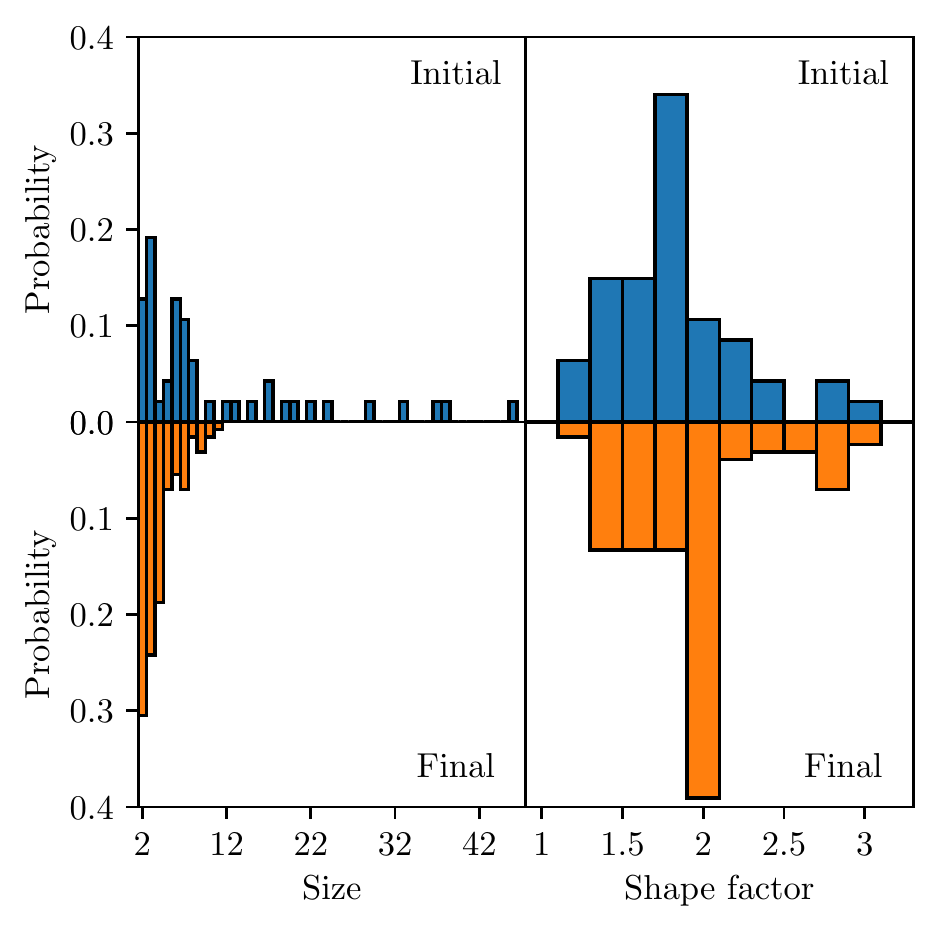}
	\setlength{\unitlength}{1cm}
	\begin{picture}(0,0)
		\put(-6.5,5.1){\includegraphics[height=2.55cm]{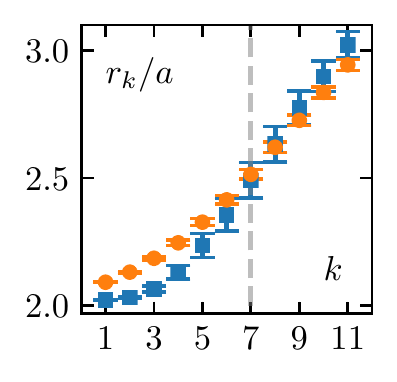}}
		\put(-6.5,1.1){\includegraphics[height=2.55cm]{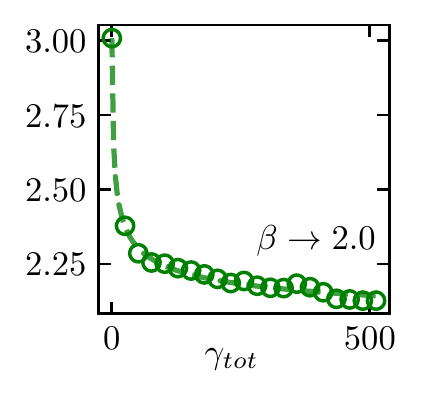}}
		\put(-6.8,7.3){(a)} \put(-6.8,1.45){(b)}
	\end{picture}
	\caption{Size and shape distributions of particle clusters in its nearest-neighbour graph 
	             under $\gamma_0=0.8$ at $\phi=40$\%.
		     (Top) Initial statistics. (Bottom) Final statistics.
		     (Inset, a) Average distance to the $k$th nearest neighbour; 
		     blue squares for the initial distances, orange circles final.
		     (Inset, b) Evolution of the number variance scaling exponent $\beta$ 
		     obtained under the same $\phi$ and $\gamma_0$, 
		     but in a much larger computational box ($N=4096$).
		     The dashed line is a fit of the data according to Eq.~\eqref{eq:beta}.}
	\label{fig:nng}
\end{figure}

Fig.~\ref{fig:nng} shows the size and shape distributions of particle clusters for a suspension under $\gamma_0=0.8$ at $\phi=40\%$.
The initial graph is composed of clusters from a wide range of sizes,
with the biggest cluster approaching 10\% of the sample size ($N=500$ in this case).
The NNG at steady state, however, is dominated by small clusters of less than 4 particles.
The dissolution of large clusters is a sign of collective organization,
implying that the distance between adjacent particles are enlarging throughout the suspension.
This can be seen directly in the inset (a) in Fig.~\ref{fig:nng},
where we plot the average distance to the $k$th nearest neighbour in the initial and final configurations.
The distance increases up to $k=7$, with the final $r_1/a=2.1$.
Correspondingly, a doublet made of two nearest spheres at $r/a=2.1$ has a shape factor of 1.93,
which explains the peak in the final distribution of the shape factor.

Lastly, the uniform increase of local separations between particles at a fixed volume implies suppression of density fluctuations at large length scales.
The latter is a hallmark of \emph{hyperuniformity},
which has been recently proposed as a distinguishable state of matter \cite{torquato2003local, torquato2018hyperuniform}.
Here, density fluctuations can be quantified by the variance of the number of particles within a randomly located sphere of radius $r$,
viz.~$\sigma_n^2(r) \equiv \langle n(r)^2 \rangle - \langle n(r) \rangle^2$,
for $r \in [2a,L/4]$, where $L$ is the length of the periodic box.
Then, the exponent of the scaling law $\sigma_n^2(r) \sim r^\beta$ measures the degree of hyperuniformity:
in $d$-dimensional space, $\beta=d$ corresponds to non-hyperuniform structures (e.g.~ordinary fluids),
whereas materials with $\beta \in [d-1,d)$ are considered hyperuniform 
(e.g.~crystals have $\beta=d-1$; other forms of scaling are also possible \cite{torquato2018hyperuniform}).
We compute $\beta$ for the same parameters $\gamma_0=0.8$ and $\phi=40\%$, 
but without far-field hydrodynamics in a much larger computational box ($N=4096$), 
as shown in the inset (b) in Fig.~\ref{fig:nng}.
The data plotted in total strain $\gamma_\text{tot}$ is well-fitted by
\begin{equation} \label{eq:beta}
    \beta = (\beta_0-\beta_\infty)(1+\gamma_\text{tot})^\nu + \beta_\infty,
\end{equation}
where $\beta_0$ and $\beta_\infty$ are the initial and steady-state values of $\beta$, respectively.
The fitting yields $\nu=-0.29$ and $\beta_\infty =2.0$,
strongly suggesting that system will eventually become hyperuniform.
We note that the size of our largest simulation  ($N=4096$, $L/4=8.8a$)
may still be too small to conclude on the large-scale structural correlations.
Nevertheless, the different characteristics of the suspension microstructure are consistent 
and in agreement with previous predictions and experiments \cite{torquato2018hyperuniform}.

\section{Concluding remarks}

In summary, we have presented a numerical study of periodically sheared non-colloidal suspensions
under various oscillation amplitudes and at different volume fractions.
The main finding is a generic rheology 
\re{characterized by a reduced viscosity, an enhanced intracycle shear thinning, 
the onset of a finite second normal stress difference as well as its frequency doubling within a cycle,}
all near a volume fraction dependent oscillation amplitude.
This rheological behavior is closely related to the emergent dynamics of suspensions under cyclic shear.
In fact, we show that the amplitude inferred from rheological measures also predicts the reversible-irreversible transition.
These findings are explained based on, firstly, the stress asymmetry caused by flow reversal and,
secondly, the shear-induced diffusion.
Although the basic phenomena have been studied separately in the past,
the connection between them in the context of RIT was unclear.
This paper was motivated to bridge the gap.

Finally, despite our focus on non-colloidal suspensions of frictionless hard spheres of equal size,
the general physical picture depicted here should hold in a broader context.
As long as the structural organization is dominated by volume exclusions,
there could always be a configuration of least density fluctuation, 
thus the specific correspondence between rheology and emergent dynamics,
regardless of the particle shape, size distribution, surface roughness, and so on.
Furthermore, evolution towards a state of least density fluctuation \re{(or crystallization at high packing fractions \cite{ghosh2022coupled})} seems favorable, 
since the microstructure associated with it reduces the energy dissipation under shear.
At present, there is a growing interest in understanding the collective dynamics of \emph{active suspensions}
that are made of self-propelled particles such as bacteria or algae,
where long-range hydrodynamic interactions are of greater importance \cite{elfring2021active}.
Extending the current study in those settings may lead to fruitful future research.

\begin{acknowledgments}
The research is sponsored by Canada’s Department of National Defence through Contract No.~CFPMN1-026.
\re{Z.G.~acknowledges support from the Swedish Research Council (grant No.~2021-06669VR) and thanks S.~Bagheri for hospitality.
Finally, the authors wish to thank S.~Torquato, P.M.~Chaikin, V.~Shaik, M.~Boudina, and the anonymous referees for useful comments.}
\end{acknowledgments}


\bibliographystyle{apsrev4-1}
\bibliography{main}
\end{document}